# 自组装半导体量子点的电子学性质研究进展

孙 捷*　金 鹏　王占国

(中国科学院半导体研究所半导体材料科学重点实验室, 北京 100083)

**摘要：** 自组装半导体量子点是人工设计、生长的一种具有量子尺寸效应、量子干涉效应、表面效应、量子隧穿和库仑阻塞效应以及非线性光学效应的新型功能材料。该材料由于晶体缺陷少、材料制备工艺相对简单等优点而在未来纳米电子器件的研制中有重要的应用价值。本文按照纵向输运、横向输运、电荷存储的顺序，扼要评述了自组装半导体量子点电子学性质的最新研究进展，并对目前存在的问题和发展前景作了分析。

**关　键　词：** 自组装半导体量子点；　量子尺寸效应；　纳米电子器件

**中图分类号：** TN304

*联系人。电话：010-82304274。传真：010-82305052。电邮：albertjefferson@sohu.com.

## Progress in the Research of Electronic Properties of Self-Assembled Semiconductor Quantum Dots

SUN Jie*, JIN Peng, WANG Zhan-Guo

(Key Laboratory of Semiconductor Materials Science, Institute of Semiconductors, Chinese Academy of Sciences, Beijing 100083, China)

*Sun Jie, doctorial student. Tel.: 010-82304274. Fax: 010-82305052. E-mail: albertjefferson@sohu.com.

Project supported by Special Funds for Major State Basic Research Project of China (Nos. G2000068303；2002CB311905), National high technology research and development program of China (Nos. 2002AA311070; 2002AA311170) and National Natural Science Foundation of China (Nos. 60306010；90101004),
Received August 17, 2004.

**Abstract:** Self-assembled semiconductor quantum dot is a new type of artificially designed and grown function material which exhibits quantum size effect, quantum interference effect, surface effect, quantum tunneling-Coulumb-blockade effect and nonlinear optical effect. Due to advantages like less crystal defects and relatively simpler fabrication technology, that material may be of important value in future nanoelectronic device researches. In the order of vertical transport, lateral transport and charge storage, this paper gives a brief introduction of recent advances in the electronic properties of that material and an analysis of problems and perspectives.

**作者简介**：孙捷，男，汉族，1977 年 8 月生，博士,从事量子点材料的分子束外延生长研究。

# 1 引言

纳米科学是研究纳米尺度（0.1-100nm）内原子、分子和其他类型物质运动、变化的科学，是包括物理、化学、生物、材料、电子、力学、制造等多学科的交叉体系。其中，纳米材料具有核心和基础的地位。半导体量子点作为纳米材料的重要内容，正在受到人们的极大关注。本文将扼要对自组装半导体量子点的电子学性质及相关器件研究进行评述。

# 2 自组装半导体量子点

## 2.1 量子点在纳米电子学中的应用前景

量子点也称人造原子[1, 2]，通常指人工制造的尺寸为 0.1-100nm 的小系统，其中含有 1-10000 个可控制的电子。它的尺寸比团簇[3]大，但小于当今光刻的精度。量子点在三个维度上的尺寸都与该方向电子的平均自由程可比拟甚至更小，因此载流子在三个方向上运动都受到了约束，能量本征值都是量子化的，故称为零维材料。它有与三维体材料显著不同的性质，主要包括量子尺寸效应、量子干涉效应、表面效应、量子隧穿与库仑阻塞效应和非线性光学效应等，在纳米电子与光电子器件研制方面有极重要的前景。

随着信息时代的到来，电子器件趋向于更小、更快、更冷。目前国际上已经生产最小线宽为 0.13μm 的硅集成电路，但当进一步发展到最小线宽小于100nm 即所谓纳米电子器件时，却受到光刻技术精度的限制[4]。为解决这些困难，分子电子学[5]应运而生，但人们渐渐认识到在微电子学与分子电子学之间应该有个过渡——纳米电子学[6,7]，即信息加工的功能元件是由有限个原子构成的聚集体，例如半导体量子点。电子的维度被限制得越多，它的束缚态能级就越精确，如果我们尽量缩小器件的电阻、电容，它就会有很好的开关特性。因此，零维材料量子点成为制造纳米电子器件的理想材料之一。由于它作为人工原子的特殊电子结构[8,9]，使得纳米电子器件有许多与传统器件不同的性质：（1）电导量子化，即电导为量子化的台阶值，而非经典情形的常数（电流/电压）；（2）库仑阻塞效应，即量子点的充放电过程不连续，充入一个电子所增加的能量为 $e^2/2C$，其中 C 为量子点电容。由于 C 很小，故库仑阻塞能较大，使电子只能一个一个地传输；（3）普适电导涨落和量子相干效应等。纳米电子器件的输运理论也与经典理论大不相同，比较典型的如用经典理论加量子条件求解电子器件问题的正统理论[10]等等。

## 2.2 量子点材料的制备方法

量子点的制备可采用化学方法与物理方法。利用胶体化学原理制备量子点是化学方法的典型例子，比如我们实验室用水/甲醇溶液制备的掺铕纳米晶态 ZnS 量子点[11]。目前这类工艺仍在发展之中。物理方法又可分为"自上而下"和"自下而上"两类。利用自上而下方法制备的量子点形状、分布可以控制，但容易损伤而引入缺陷和玷污。通过光刻技术（例如在量子阱或超晶格结构的基础上用高分辨电子束曝光直写刻蚀[12]）和量子阱界面波动（例如利用量子阱势能波动在低温下使两维激子局域化[13]）制备量子点属于自上而下的办法。利用晶体生长的 S-K(Stranski-Krastanow)模式进行应变原位自组装生长半导体量子点属于自下而上的方法，也是最简便、最成熟的技术，我们实验室与国际同步开展了此类研究[14]。它的原理是：晶格失配度适中的两种材料，例如 Ge/Si, InAs/GaAs 等，在分子束外延（MBE）或金属有机化学气相淀积（MOCVD）初始阶段是二维平面生长，随着厚度的增加产生应变积累，导致在临界厚度时外延层由层状生长转变为三维岛状生长以便降低系统能量，最终形成了均匀分布且无位错的量子点。

通过优化生长条件，可使量子点尺寸和分布的不均匀性≤10%，密度控制在 $10^8$-$10^{11}cm^{-2}$。

应变自组装量子点虽然制备工艺简单、点的尺寸小且无损伤，但由于浸润层上的成核是无序的，故其形状、尺寸、分布均匀性和密度难以控制。为了充分发挥自组装量子点在器件研制中的作用，需要对其生长进行人工调控[15]。目前人们主要通过对材料体系的应变分布设计（晶格失配度和衬底晶向的选择、缓冲层与盖层的设计等）、生长动力学（生长温度、生长速率等）控制和生长工艺优化来控制量子点的自组装生长。例如，在 GaAs 衬底上生长 $In_xGa_{1-x}As$ 量子点时，可通过改变 In 的组分和采用高指数面衬底进行应变分布设计。我们实验室[15]系统研究了在不同取向晶面上和在（311）B 面上不同 In 组分下生长的 $In_xGa_{1-x}As$ 量子点的表面形貌，结果显示当 In 的组分为 0.4 时，在（311）BGaAs 衬底上制备的 $In_xGa_{1-x}As$ 量子点的均匀性和有序性最好。S-K 模式中，量子点的形成是个动力学限制过程。如果在一定条件下进行生长动力学控制和工艺优化，可使量子点尺寸自发地均匀化（即尺寸自限制效应 SSL）[16]。文献[16]中给出了在 $3\times 10^{-7}$Torr 砷压下淀积速率为 0.035ML/s 时生长的尺寸比较均一的 InAs/GaAs 量子点形貌图。这种 SSL 效应源自量子点形成过程中二维平台边缘的压应力和三维岛的四个 (136) 晶面。总之，人们提高 S-K 量子点均匀性和有序性的研究仍在进行之中。

## 3. 纵向输运

共振隧穿二极管(RTD)是基于量子隧穿现象的一种负阻器件，具有高速、高频、低压、低能耗等优点，因而成为下一代集成电路的发展方向之一[17]。由于 RTD 有诸如负微分电阻的复杂输运特性，故可实现更精确的逻辑功能，或在相同情况下使集成电路简化[18]。早在 1974 年 Esaki[19] 等人就观测到了半导体量子阱双势垒共振隧穿现象，目前它已可与 CMOS 器件混合集成[20]。常见的 RTD 为双势垒结构，势垒间为低维半导体有源区，势垒外为发射极和集电极。施加一定偏压时，能带发生倾斜，原本较低的发射极电子能级与低维结构电子基态能级一致，电子便有较大概率共振

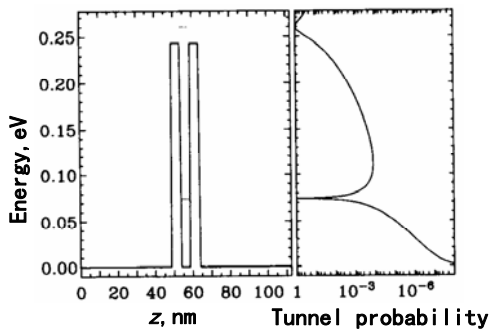

图 1 量子阱 RTD 的理想能带图（左图）。右图给出了不同能量入射电子的量子隧穿几率。
Fig. 1 Ideal energy band profile of the quantum well RTD (left). The right figure shows quantum tunneling coefficients of electrons with different incident energies.

隧穿通过双势垒。继续增大偏压使有源区电子基态能级低于发射极能级时，共振隧穿截止，出现负微分电阻现象。随偏压的升高，电子还可通过低维结构中激发态能级发生共振，使电流-电压曲线呈现振荡或台阶特性。图 1 给出了在理想化的 GaAs/Al$_{0.3}$Ga$_{0.7}$As 对称双势垒结构中（左图，假设导带带阶为能隙差的 65%）计算出的不同能量发射极电子的量子隧穿几率（右图）[21]。显然，在大多数能量下电子将被势垒反射，而能量与量子阱束缚态能级一致时，电子的透过率可达 100%。在传统量子阱 RTD 中，有源区电子能级实际上是一系列能量子带。而对于量子点 RTD，由于电子能级十分准确，所以理论预期有更精细的 I-V 特性。这类量子点共振隧穿器件是典型的基于电子在平行量子点生长方向上的输运特性制成的纵向输运器件，近年来已成为科研的热点[22-25]。为消除自组装量子点尺寸波动导致的电子能级展宽效应，日本东京大学 Sakaki 小组[26]研制了单量子点 RTD。最近，我国科研人员[27,28]也开展了对量子点 RTD 的研究。

瑞典隆德大学 Samuelson 小组[29-31]研制的一种量子点 RTD 的有源区理想能带示于图 2[31]。由于自组装量子点的大小、形状尚不是非常均一，若器件有源区量子点数目过大，这种不一致性将使其电学性能变差。为减小器件中点的数目，必须降低其生长密度[32,33]。利用 InP/InAs 界面化学反应（气相 As 原子迅速置换最外层 P 原子），低压 MOCVD 生长的 InAs/InP 量子点的面密度可减至 $10^6$cm$^{-2}$ 量级[31]，图 2 中的量子点（衬底是重掺杂 InP）即用这种方法制成，有源区结构由下至上依次为：5nmInP 势垒、InAs 量子点、15nmInP 势垒、InAs 量子点和 12nmInP 势垒。该器件采用三势垒结构，有源区两侧为 InGaAs 间隔层、渐变掺杂的 InGaAs 发射极与收集极和欧姆接触，电子在偏压

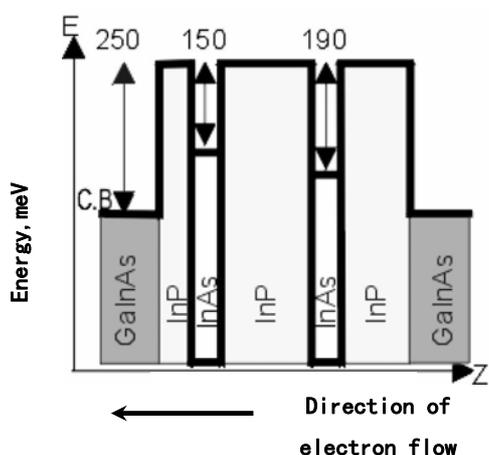

图 2 量子点 RTD 有源区的能带示意图
Fig. 2 Energy band diagram of the QD-RTD's active region.

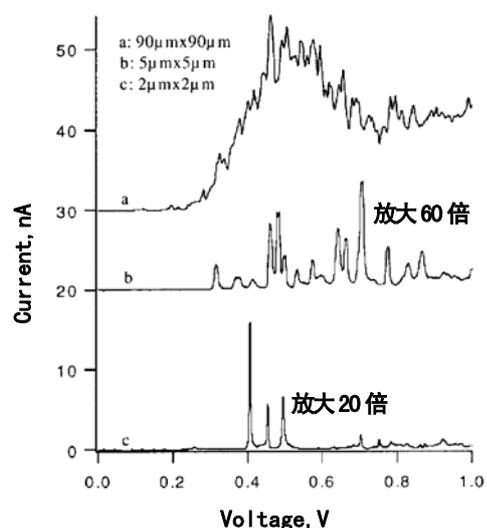

图 3 具有不同大小台面的量子点 RTD 的电流-电压特性
Fig. 3 I-V curves of QD-RTDs with assorted-sized mesas.

作用下由顶层隧穿入衬底。由于应力场的作用，下层点（5nm 高，30nm 宽）与上层点（6nm 高，40nm 宽）是耦合的，也称人造分子，其估算的电子能级可见图2。器件中异质结构的设计使得在适当偏压下两层量子点中能级恰好均与发射极电子能级对准，以产生尖锐的共振隧穿电流峰。虽然量子点密度已经很小，但为进一步减小其不均匀性带来的负面影响，需要在器件表面腐蚀出不同大小的微台面。图3[29]的 I-V 特性中，a 曲线对应于90μm×90μm 台面，器件中含千百个量子点；b 曲线(5μm×5μm 台面)的隧穿电流峰就要明显一些；而对于 c 情形，2μm×2μm 的欧姆接触仅能控制一个或几个量子点，从而在 4K 下可获得高达 1300 的峰谷比和仅 6mV 的半高宽，为迄今此领域的最好结果。

## 4. 横向输运

随着集成电路工艺的迅猛发展，其基本元件场效应管的特征尺寸必然进入纳米量级，器件将脱离经典原理而表现出量子性质。在纳米加工学基础上，人们已经开始研究全新的单电子晶体管（SET），它与传统晶体管的差别有如滴定阀门与水龙头的差别。拧水管可以对水量调控（栅极控制），但却难以进行滴定管那样的细调。由于库仑阻塞效应，电子只能逐个通过器件，呈现单电子（或准单电子）输运行为。最近，用半导体或金属量子点研制的单电子晶体管已经有 10meV 以上的电子增加能（库仑阻塞效应中，库仑充电能和新增电子的量子动能之和称为电子增加能，即 electron addition energy，是衡量充电效应能量范围的通用参数），有的甚至超过 100meV，在室温下可观测显著的库仑阻塞振荡特性[34-36]。利用自组装 Si 量子点，韩国高丽大学 Huang 小组[37, 38]研制了在室温下有库仑阻塞效应的晶体管[37]。首先在 p 型 Si 上热生长 300nmSiO$_2$，然后利用低压化学气相淀积技术在 620°C 下自组装 Si 量子点[39]。三个纳米 Al 电极(源 S、漏 D、旁栅 G)是通过标准电子束光刻和剥离（liftoff）工艺制成的，图4[37]中给出了其 AFM 照片。

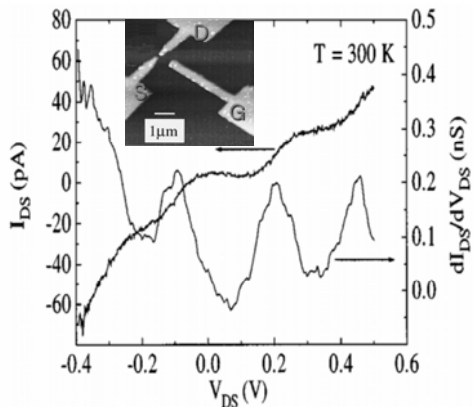

图 4 室温下量子点晶体管的 I$_{DS}$-V$_{DS}$（左）和 dI$_{DS}$/dV$_{DS}$-V$_{DS}$（右）曲线。插图是淀积的几个铝电极的原子力显微像。

Fig. 4 Room temperature I$_{DS}$-V$_{DS}$ (left) and dI$_{DS}$/dV$_{DS}$-V$_{DS}$ (right) curves of the QD transistor. The inset is AFM image of as-deposited Al electrodes.

电子通过源漏间的量子点进行输运,衬底背面还制做了背栅极。虽然源漏间量子点数有可能大于1(两电极间距30nm,约为自组装Si点平均间距16nm的二倍),但研究表明点尺寸较均一且结电容较小时仍能观测到库仑台阶[40]。在图4的室温$I_{DS}$-$V_{DS}$和$dI_{DS}/dV_{DS}$-$V_{DS}$(微分电导-电压)图线中,周期0.27V的库仑振荡清晰可见,据此可推算出量子点总结电容$C_D+C_S$约为0.296aF。在源-漏电流随旁栅压和背栅压变化的曲线中亦可发现类似的振荡或台阶,并由此得知耦合的旁栅电容、背栅电容分别为0.285aF和1.01aF。因此,点的总电容为1.59aF,对应于7.3nm的量子点平均直径,而这与扫描电镜形貌结果自洽。这种器件可在常温下观察到电子通过源漏间隧穿结输运时的库仑阻塞效应,是未来纳米集成电路基本元件的前身。

还有一些基于量子点横向输运性质的基础研究往往采用量子点和量子阱耦合的结构[41-43]。在这些器件中,通过量子点中载流子充放电对量子阱沟道中二维电子气(2DEG)输运的调制作用可以获得预期的电流-电压特性(振荡、台阶等)。英国剑桥大学卡文迪许实验室[44,45]研制的一种利用量子点对调制掺杂场效应管进行浮栅(floating gate)调控的单光子探测器的结构可见图5[44]。自组装量子点有源区与GaAs量子阱沟道耦合,中间隔有薄$Al_{0.33}Ga_{0.67}As$势垒。源和漏电极是AuGeNi欧姆接触,栅极是用7nmNiCr制成的半透明肖脱基接触。由于量子点电子基态能级低于GaAs沟道的导带边,所以每个点都能俘获几个电子,相当于对量子点进行了负性充电。这些过剩负电荷将产生排斥势,使沟道内接近量子点的区域电子耗尽,造成相对较低的电子迁移率,引起电子气电导下降。2DEG的电

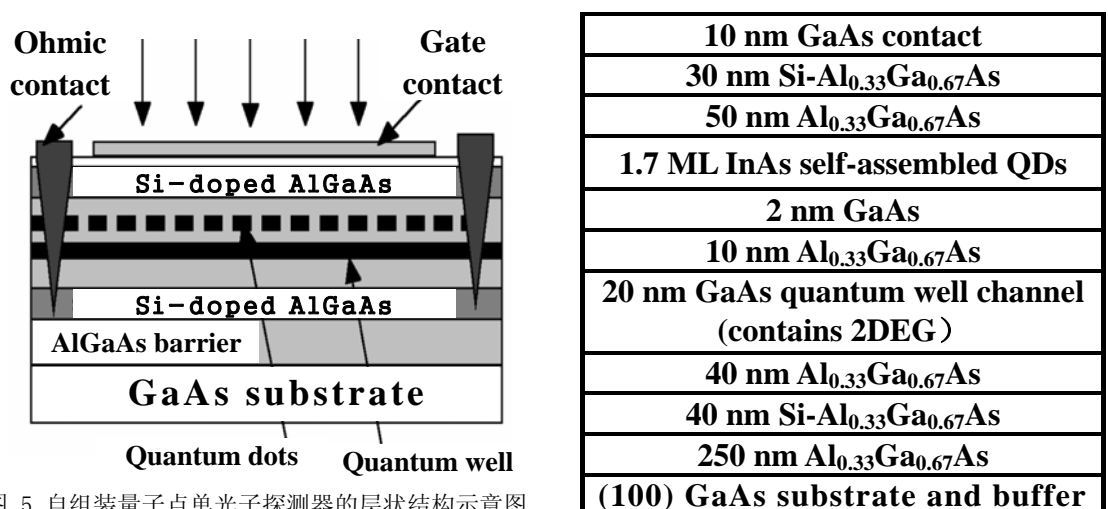

图5 自组装量子点单光子探测器的层状结构示意图
Fig. 5 Diagramatic sketch of epitaxial layers of self-assembled QD single photon detector.

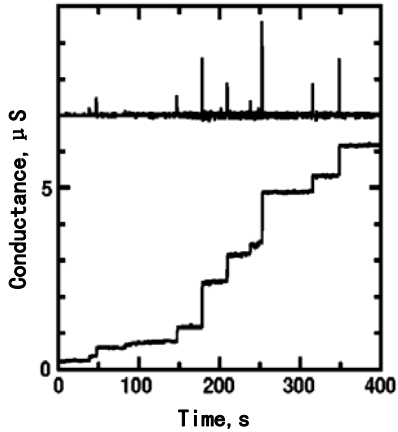

图 6 单光子探测器的源-漏电导-时间曲线（下部）；上部为时间微分电导图线

Fig. 6 Single photon detector's S-D conductance as a function of time (lower part). The upper part is the time differential of conductance.

导较低时，沟道电流对量子点的过剩电荷极其敏感。此时若用光束透过栅极照射，将会在器件中产生电子空穴对，量子阱内的光生空穴在内电场作用下将隧穿入量子点并与过剩电子复合。同时，阱内的光生电子则保留在 2DEG 中。因此，光照可以减少量子点内过剩电子数目（相当于对量子点放电）和增加两维电子气密度，从而减轻量子点负电性对沟道电流的限制作用，使测量电导值上升。若有源区量子点数目足够少，我们就可检测出一个量子点俘获单个光生载流子引起的 2DEG 电导变化，从而进行单光子探测。图 6[45]显示了在工作电流为 2μA 的发光二极管(LED)照射下 2DEG 电导随时间的变化情况。进行单光子探测前先加 10s 0.76V 栅压对量子点充电，以创造源−漏低电导的初始条件。利用一个大面积光电二极管检测到的光生电流推算，照射到场效应管栅极上的光子束流约 7.5 光子/秒。图 6 中电导图线有一系列方阶，每次阶跃都是由单光子辐射引发一个量子点放电所致。虽然这种单光子探测器量子效率仅为 0.48％，尚远低于传统方法[46]，但毕竟是一种全新的探索。导致低探测率的原因主要是大多数光子直接透过量子阱被衬底吸收，今后可用增加厚的吸收层和提高栅极材料透明度方法加以优化。

## 5. 电荷存储

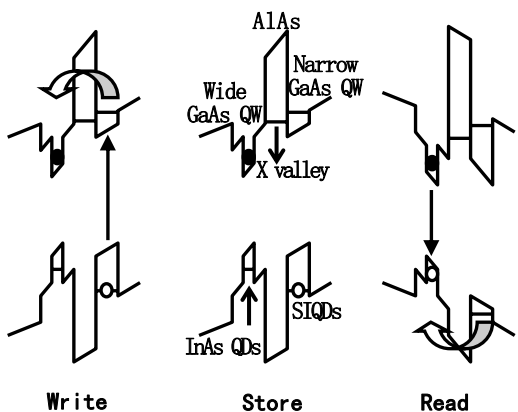

图 7 量子点激子存储器的读写过程（能带示意图只画出了有源区）

Fig. 7 Writing and reading procedures of QD exciton memory (only the active region is shown).

信息存储容量和速度是计算机功能的重要指数，目前存储一比特(bit)信息需移动至少十万个电子[47]，而以量子点为媒介实现同样功能只需对有限个甚至单电子操作。由于工作的电子数大大减少，信息存储密度、器件功耗和响应速度都会明显提高。这方面的基础和应用研究[48-50]显示，自组装半导体量子点是理想的新一

代存储介质之一。依据量子点充放电时电子流动方向，此类器件亦可归入横向或纵向器件，但在此将对其单独评述，以突显其重要性。

著名量子点专家，美国加州大学圣巴巴拉分校 Petroff[51]领导的小组[52-54]开发的量子点激子存储器的有源区能带结构示意于图 7[52]。他们用 MBE 在半绝缘 GaAs 衬底上生长了 40 周期 AlAs/GaAs(20Å/20Å)超晶格（为使外延层平整）和 1000Å $n^+$GaAs($10^{18}cm^{-3}$)背电极接触层，并在接下来的 400Å$Al_{0.5}Ga_{0.5}$As 势垒上依次生长了 68 Å GaAs 量子阱（阱中插入了自组装 InAs 量子点(QD)层）、60Å 或 100Å AlAs 势垒、25Å GaAs 量子阱、500Å $Al_{0.5}Ga_{0.5}$As 势垒和 7 周期 AlAs/GaAs(20Å/10Å)超晶格，最后在顶部 50Å GaAs 盖层上制备了半透明肖脱基栅极。下层较厚的量子阱中，InAs 量子点与 GaAs7％晶格失配产生的应力场可透过 AlAs 势垒传递到上层较薄的 GaAs 量子阱中，形成掩埋的与下层自组装量子点(QD)耦合的应力诱导量子点(strain-induced quantum dot, 即 SIQD)。设计的薄量子阱厚度使其电子基态能级稍高于 AlAs 势垒中 X 能谷，这种薄阱电子基态能级－AlAsX 能谷－厚阱中量子点电子能级的阶梯式结构有利于 SIQD 中的光生电子在几十皮秒内弛豫到 QD 中（施加负偏压效果更佳）。在这种耦合量子点结构中，写入激光脉冲产生的电子和空穴被分别存入 QD 和 SIQD 内，加一定偏压时，SIQD 中空穴隧穿入 QD 并与电子复合产生读出光，这就是存储器的工作原理。实验中，观测到 QD 在激光写入时 1.25eV 的光致发光峰后，延迟 3s 加 0.5V 偏压，确实有较强的 1.25eV 读出光脉冲出现，标志着光生激子在 QD 中成功复合。3K 时 0.5V 偏压下读出光的信号衰减可见图 8[52]。衰减曲线可用两个指数函数 $4×10^{-2}e^{-0.57t}$ 和 $3×10^{-2}e^{-0.18t}$ 拟合（分别对应于 5μs＜t＜800ms 和 800ms≤t＜3s，其中 t 为存储时间（秒））。信号衰减机制主要有 3 种：(1)QD 中电子与 SIQD 中空穴由于波函数交叠而间接复合；(2)载流子被 AlAs 势垒中的深能级俘获；(3)电

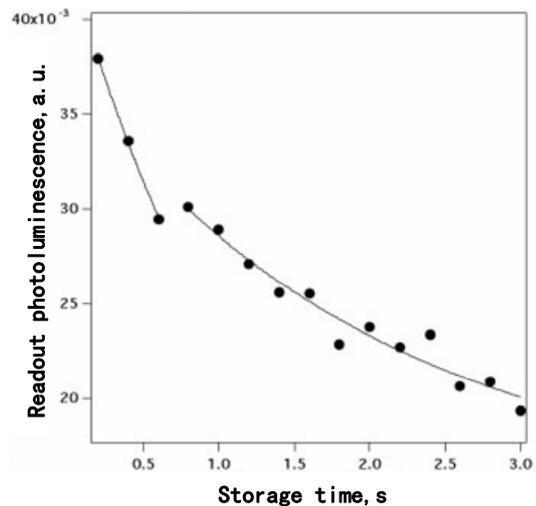

图 8 低温下读出信号随时间的衰减图线（积分强度）

Fig. 8 Low temperature attenuation curve of the readout signal as a function of time (integrated intensity).

子与 MBE 中碳背景受主产生的空穴复合。尽管使信号衰减的因素相当繁杂，但这种通过双层耦合量子点使载流子局域化的结构依然有 3s 的信号半衰期，存储时间长达 10s，比量子阱存储器[55]高一两个数量级。

虽然基于持久谱线烧孔(persistent spectral hole-burning)效应的频率选择光存储概念早已问世[56]并在有机高聚物和玻璃态物质中有深入研究[57]，但其在量子点电荷存储方面的应用却始于日本富士通公司近年来的工作[58,59]。当前，自组装量子点的形状、体积尚不能被很好地控制，由生长方向上尺寸波动导致的光谱不均匀性线宽 $\Gamma_i$ 可达 80－100meV[60]，对于制备量子点光电器件往往不利，但我们却恰好可利用这一点研制谱线烧孔存储器。一束能量为 hν 的光子入射时，本征吸收长波限（与量子点大小密切相关）恰为此值的量子点中将产生电子－空穴对。由于它们是费米子，对随后的光吸收过程有阻碍作用[61]（即相应能态被载流子全部占据），所以在吸收谱或本征光电导的光谱分布曲线的相应波长处将有烧孔现象。设法使电子或空穴二者之一隧穿出量子点可降低其复合率，亦即提高了谱线烧孔的寿命。如果将烧孔与否用(1,0)表示，那么在谱线不同波段处烧出 n 个可分辨的孔，就能以波长为域存储 n 个比特（bit）。显然，不均匀性线宽与均匀性线宽之比 $\Gamma_i/\Gamma_h$ 是决定存储密度的重要因素。理论上，δ 函数状的能态密度使量子点的 $\Gamma_h$ 很小，因此其信息存储密度高于传统谱线烧孔介质，并有望比目前光盘存储密度提高几个数量级[59]。文献[58]给出了富士通公司开发的一种含五层堆垛自组装 InAs 量子点（密度 $5\times 10^{10}cm^{-2}$，相邻两层间隔 20nmGaAs）的 pin 光电二极管结构图。图 9[58]是 5K 时在可调谐连续波长钛蓝宝石(TiS)激光器照射下该器件光电流的光谱分布图。1064nm 连续波长钇铝石榴石(YAG)激光器也被用来照射器件以进行谱线烧孔。可以看到，当光电二极管外加偏压超过 4V 时有明显的谱线烧孔效应，孔宽小于 1nm，系由 8mWYAG 激光器的几条激射线（箭头所示）烧孔叠加而成。一定大小的量子点中由于吸收了 1064nm 波长激光而产生大量电子－空穴对，其中电子因

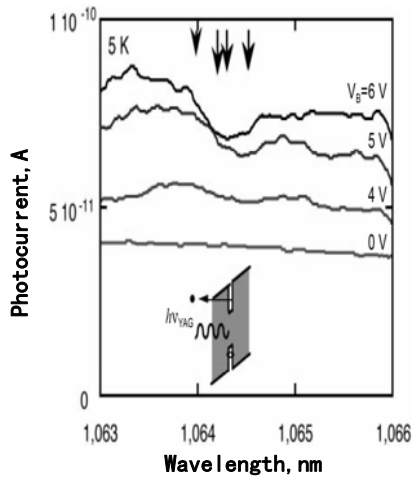

图 9 光电流的光谱分布图中的烧孔效应
Fig. 9 Hole burning effect in the spectrum distribution curve of photocurrent.

为有效质量较小而在偏压下更易在复合前隧穿出量子点(见图9中的能带图)，这样剩余的空穴就会阻碍对该波段TiS激光的吸收而形成烧孔。这种二极管的烧孔寿命现在只有$1.2\times10^{-6}s$，在室温下应用尚不现实，但却是利用自组装半导体量子点的不均匀性研制电荷存储器件的良好范例。

## 6. 问题与展望

十几年来，自组装半导体量子点的光学应用已有长足进步[62,63]，但其电学性质的研究尚在起步阶段。目前的困难主要集中于三个方面。首先，人们尚不能精确控制量子点的形状、尺寸、面密度和体密度。除一些方面（如谱线烧孔或超辐射LED）外，大多数情形下量子点的不均一性对器件研制是不利的。自组装量子点的分布受表面台阶性质、原子迁移、化学组分和应力分布等诸多因素的综合作用，表现出复杂的行为，需要实验研究、计算机模拟和理论分析的密切配合才能透彻了解。采用图形化衬底和不同取向晶面选择生长技术可使量子点均匀性有一定提高，但仍未达到令人满意的程度。其次，量子点生长技术较成熟的Ⅲ-Ⅴ族化合物半导体虽有较高电子迁移率和较小的电子有效质量，但由于其表面态密度较大和无良好的隔离介质膜而并非最理想的纳米电子材料。硅基半导体（如GeSi/Si）[64]兼具低界面态密度和硅高纯、完整的优点，若能解决介质隔离问题，有望成为首选材料之一。新兴的碳纳米管[65]、有机电子材料[66]发展势头迅猛，但与应用尚有相当距离。第三，现有纳米加工技术还不能满足器件制备的需要。纳米电子器件的大规模集成（$10^9$-$10^{10}$/cm$^2$）要求快速、廉价、精确地对它们进行加工和连接。如果按1μm左右线宽的工艺水平，电子器件必须工作在液氦温度（4.2K）；若要实现液氮温度（77K）工作，器件尺寸（以Ⅲ-Ⅴ族材料为例）须在50nm以下。由此可见，发展纳米级精确、高速和无损的加工工艺与相应装置（如扫描探针显微镜（SPM）、微机电系统（MEMS）等）是实现量子点纳米电子器件产业化的主要难点之一。

20世纪到21世纪是电子器件特征尺寸由微米向纳米过渡的时期。著名电子学家Chiabrera[67]从理论上指出了微电子元件的物理极限，美国Science杂志更是预测了新一代纳米计算机[68]。全球各国均将纳米研究作为科技竞争的主要领域，最近英特尔公司已开发出世界最小的晶体管[69]。三十多年来，我国在集成电路领域一直处于被动地位，但面临新兴的量子点纳米电子学，我们却与西方基本

处于同一起跑线上。基于应变自组装半导体量子点的纳米电子器件性能优越，应用背景明确，它的研制必将使我国在一个高起点上参与国际竞争，取得主动地位。

**参考文献**